\documentstyle[aps,prl]{revtex}

\draft

\begin{document}

\twocolumn[\hsize\textwidth\columnwidth
\hsize\csname@twocolumnfalse\endcsname\draft\pagenumbering{roma}
\title {\Large\bf Possible composite-fermion liquid as a crossover from Wigner crystal to bubble phase in higher Landau level}
\author{Shi-Jie Yang, Yue Yu and Zhao-Bin Su}
\address{Institute of Theoretical Physics, 
Chinese Academy of Sciences, P. O. Box 2735, Beijing 100080, China}
\maketitle

\vspace{15mm}
\begin{abstract}
The ground state cohesive energy per electron of the composite 
fermion (CF) Fermi sea, the Laughlin state and the charge density 
wave (CDW) at higher Landau levels (LLs) are computed.
It is shown that whereas for $n\geq 2$ LL, the CDW state is generally 
more energetically preferable than those of the CF liquid and 
the Laughlin liquid, the $\nu =4+1/6$ CF liquid state
unexpectedly has lower ground state energy than that of the CDW state. 
We suggest this CF liquid between the Wigner crystal and the bubble 
phase may lead to the crossover from the normal integer quantum Hall
liquid to the novel re-entrant integer quantum Hall state observed in the
recent magneto-transport experiments. 
\end{abstract}
\vspace{4mm}
\pacs{PACS number(s): 73.20.Dx, 73.40.Hm, 73.40.Kp, 73.50.Jt}
\vskip 2pc
] \newpage \pagenumbering {arabic}

\indent While fractional quantum Hall effects (FQHE) are impressive for
their odd-denominator fillings, the remarkable phenomena for
even-denominator fillings have caused great interests in the last decade
\cite{book1}. By using the concept of CF \cite{Jain1,book2},
Halperin {\it et al.} suggested that the $\nu=1/2$
system can be viewed as a spin-polarized Fermi liquid of CF \cite{Halperin}.
The only incompressible state at even-denominator filling $(\nu=5/2)$ in the
single-layer 2-dimensional electron gas  is now widely accepted
as the pairing of CF in a vanishing effective magnetic field \cite
{Greiter,Morf1,Park}. For higher LLs, recent magneto-transport experiments
on high mobility samples in GaAs/AlGaAs heterostructures revealed new
classes of correlated many-electron states \cite{Eisenstein1}. 
The most prominent findings are the discoveries of the giant anisotropy in the
resistivity near half filling of the topmost LL\cite{Lilly,Du} and the
observation of re-entrant integral quantum Hall (RIQH) states in the flanks
of these same levels \cite{Cooper} as well as the substantial non-linearity
of the resistivity. It is considered that the highly anisotropic transport
is related to the formation of the unidirectional charge density wave (UCDW)
state, {\it i.e.}, the stripe phase \cite{Jungwirth,Phillips}. Specifically,
the state of this system may be classified by their symmetries, which are
highly analogous to those of liquid crystals \cite{Fradkin}. The possible
states include stripe crystals \cite{Fertig}, smectic, and nematic phases
\cite{Oppen,Yi,Emery,MacDonald1}. The RIQH effect was thought to be the
depinning and sliding of the Wigner crystal (WC) and reformation of the bubble phase. But no further discussions were given on the details of this transition. \newline
\indent In the present work, we carry out systematic calculations of the cohesive
energies of various competitive ground states in the lowest as well as 
higher LLs. We find that while in the lowest and second LLs the CF
and the Laughlin liquids prevail for $\nu_{n}<1/7$ ($\nu_{n}$ is the filling
factor at the $n$-th LL), the CDW
generally has a lower ground energy than those of the corresponding CF
or Laughlin liquid for LL indices $n\ge 2$. In particular, the
stripe phase dominates around $\nu_{n}=1/2 (n\geq 2)$, which is in agreement with
previous calculations \cite{Koulakov,Moessner,Rezayi}. However, we find an interesting
exception occurs at $\nu_{2}=1/6$, where the CF liquid has a lower cohesive
energy than that of the CDW phase. This result may lead to an important
experimental phenomenon. Our calculation shows that the CDW has one electron
in each bubble (the WC phase) for $\nu_{2}<1/6$ while two electrons for $
\nu_{2} >1/6$ (the bubble phase). We relate this result to the recent
observed phenomenon of the RIQH, {\it i.e.}, the metallic CF liquid phase may
appear as an intermediate state between the WC phase and the bubble phase.
We suggest that the WC first melts into the CF liquid and then the system
recrystalizes into the bubble phase as the filling at the topmost LL
increases. \newline
\indent The fractional quantum Hall states at higher LLs were first
suggested by MacDonald {\it et al}\cite{MacDonald2} by raising the Laughlin
wavefunction to higher LLs. The state wave function at the $n$-th LL is
defined as follows:\newline
\begin{equation}
\vert \Psi^{n}_{L}\rangle=\prod\limits_{i}\frac{(a_{i}^{\dag})^n} {\sqrt{n!}}
\vert \Psi^{0}_{L}\rangle
\end{equation}
Here $a^{\dag}_{i}$ is the inter-LL ladder operator, promoting the $i$-th
elestron to the next LL, and $\vert \Psi^{0}_{L}\rangle$ is the Laughlin
state in the lowest LL. It should be noted that this state in the higher LL
is in fact not the realistic electronic state because the lower LLs are
empty. The effect of electrons in the lower filled LLs was treated by
Aleiner and Glazman \cite{Aleiner} and other authors \cite{Wangzq} by
integrating out the electron degrees of freedom in the lower LLs, which
leads to a renormalization of the Coulomb interaction between electrons at
the topmost LL. \newline
\indent The correlation energy per electron can be calculated using the
density-density correlation function in the higher LL: \newline
\begin{equation}
h_{n}({\bf r})\equiv \frac{\langle \rho_{n}({\bf r})
\rho_{n}(0)\rangle-\langle\rho_{n}\rangle ^{2}} {\langle \rho_{n}\rangle}
\end{equation}
where $\rho_{n}({\bf r})$ is the projection of the density operator onto the 
$n$-th LL. This can be most effectively done in the Fourier space because $
h_{n}(q)$ is very simply related to $h_{0}(q)$ (the correlation for $n=0$)
\cite{Fogler}: \newline
\begin{equation}
h_{n}(q)=h_{0}(q) \left[L_{n}\left(\frac{q^{2}l_{0}^{2}} {2}%
\right)\right]^{2}
\end{equation}
where $L_{n}(\frac{q^{2}l_{0}^{2}}{2})$ is the Laguerre polynomial. The
correlation energy can be written as: 
\begin{equation}
E_{cor}=\frac{1}{2} \int \frac{d^{2}q}{(2\pi)^{2}} \frac{v(q)}{\varepsilon(q)%
} h_{n}(q)
\end{equation}
\indent The cohesive energy per electron is then, according to the
definition 
\begin{equation}
E_{coh}=E_{cor}-E^{UEL}
\end{equation}
where 
\begin{equation}
E^{UEL}=-\frac{\nu_{n}}{2} \int dq \frac{e^{2}}{\kappa l_{0}} \frac{%
F^{2}_{nn}(q)}{\varepsilon (q)}
\end{equation}
is the interaction energy per particle in the uncorrelated electron liquid
formed at high temperature. Here $F_{nn}(q)=L_{n}(\frac{q^{2}l_{0}^{2}}{2}
)e^{-q^{2}l_{0}^{2}/4}$ and $\epsilon (q)$ is the dielectric function which
accounts for the screenings of the Coulomb interaction between electrons at
the topmost LL by electrons in the lower LLs. Make use of the analytical
expressions of the pair-distribution functions for $\nu=1/3$ and $1/5$ in the
lowest LL obtained earlier by MacDonald {\it et al}\cite{MacDonald2}. We
reproduce the pair-distribution functions of the Laughlin states for $
\nu_{n}=1/3$ and $1/5$ in the $n=0, 1, 2$ LLs (Fig.1). It shows that the
electrons have a larger probability to approach each other in higher LLs
than in lower LLs.\newline
\indent The trial wave function for the CF liquid ground state at higher LL
can also be obtained in the same way described above \cite{Morf2} if the 
Rezayi-Read wave functions for the CF at the lowest LL are considered \cite{rere}.
We consider a system of $N$ electrons in the external magnetic field. Through
attaching $2p$ ($p$ integer) flux quanta to each electron, one constructs
the CF which experiences a vanishing effective magnetic field at
even-denominator fillings. The interaction energy of electrons is transmuted
into the kinetic energy of CFs. To compute the
pair-distribution function of the CF liquid, we work on a spherical geometry
in which the $N$ electrons move on the two-dimensional surface of a sphere
under the influence of a radial magnetic field $B$ originating from a
magnetic monopole of strength $Q$ at the center, which corresponds to a
total flux of $2Q\phi_{0}$, where $\phi_{0}=hc/e$ is the flux quantum. We
consider a filled shell system with $N=n^{2}$, because the ground state is a
single Slater determinant, which makes computations easier. Employing the
projection scheme proposed by Jain and Kamilla \cite{Jain2,book2}, we compute
the pair-distribution function of CFs for $N=25$ electrons in the lowest LL
by using the standard Metropolis Monte Carlo method. We carry out $5\times
10^{6}$ steps and the results are shown in Fig.2. Although the CFs hardly
interact with each other in the Fermi sea state, the electrons carrying more
flux avoid each other more steadly. The cohesive energies of CF liquids $%
E_{coh}^{CF}$ and Laughlin states $E_{coh}^{L}$ for $\nu_{n}=1/2p$ ($p$%
=1,2,3,4) are listed in Tables 1-4.\newline
\indent Due to the screenings by the lower LLs, the repulsive interaction
between any two electrons at the topmost LL as a function of the separation
between the guiding centers of their orbits abruptly drops at the distance
of two cyclotron radii. Such a "box-like" component in the interaction
potential makes the uniform distribution of the electron density at the
topmost LL unstable. The translational symmetry is spontaneously broken and
crystalline domains with filling factor equal to one and zero are formed 
\cite{Koulakov}. The cohesive energy of the CDW state in the
Hartree-Fock approximation is expressed as follows:\newline
\begin{equation}
E_{coh}=\frac{n_{L}}{2\nu_{n}} \sum\limits_{{\bf q}\not{=}0} u_{HF}(q)\vert
\Delta({\bf q})\vert ^{2}
\end{equation}
where $n_{L}=1/2\pi l_{0}^{2}$ is the density of one completely filled LL
and $\Delta (q)$ is the order parameter of the CDW.\newline
\indent For the bubble state, in the limit of a weak magnetic field a simple
quasiclassical picture can be given. In this case electrons can be viewed as
classical particles rotating in cyclotron orbits. It is shown that the
optimum number of electrons in a bubble is $\tilde{M}\simeq 3n\nu_{n}$,
which corresponds to the separation $\simeq 3R_{c}$ between nearest bubbles.
However, to calculate the cohesive energy accurately, one can not use the
quasiclassical approach and the CDW should be defined more precisely.
According to Fogler and Koulakov, \cite{Fogler} it can be shown that for the
bubble states, 
\begin{equation}
\Delta (q)\approx 2\nu_{n} A\frac{J_{1}(ql_{0}\sqrt{2M})}{ql_{0}\sqrt{2M}}
\end{equation}
which is just the Fourier transform of a uniform disk with a radius $l_{0}%
\sqrt{2M}$. Here the asymptotic formula for the Laguerre polynomials for $%
q\ll \sqrt{M}/l_{0}$ is used. The cohesive energy of the bubble state can be
calculated in the same way as it has been done for the Wigner crystal. The
results are different for each $M$. Therefore, one has to find the most
optimum $\tilde{M}$ corresponding to the lowest energy. The final results
for the stripe phase $E_{coh}^{UCDW}$ and the bubble phase $E_{coh}^{B}$ are
also listed in Tables 1-4 for comparison. \newline
\indent The cohesive energies of the ground states for the Laughlin liquid,
the CF liquid and the CDW are plotted in Fig.3. It can be seen that in the
lowest LL the Laughlin liquids and the CF liquids have lower ground energies
than those of corresponding CDW states for $\nu<1/7$. In the second LL, an
incompressible pairing state of CF is considered to be more energetically
preferable at $\nu_{1}=1/2$. The energy gap disappears when an in-plane
magnetic field is applied to the system \cite{Eisenstein2,Pan}. It was ever
interpreted as indicating the existence of substantial spin reversal in the
ground state \cite{Eisenstein2}. The early tilt field experiments, however, missed
an important point. In addition to suppressing the $\nu=5/2$ FQHE state, the
tilted field leaves the transport in the $n=1$ LL highly anisotropic.
Recently, the present authors demonstrated that the pairing of CFs is
destroyed by the in-plane field and finally transforms to the UCDW  \cite{Yu}. 
On the other hand, our calculations show that a CF liquid may exist at 
$\nu_{1}=1/4$ and $1/6$, in agreement with Morf and d'Ambrumenil \cite{Morf2}.

For the $n\geq 2$ LL, the CDW states are generally favorable. The Laughlin liquid
and the CF liquid are ruled out in the region of $\nu_{n}< 1/6$, and specifically, the
stripe phase dominates around $\nu_{n}=1/2$. This result agrees with
earlier works using the Hatree-Fock method and recent numerical studies
carried out by Rezayi, Haldane and Yang \cite{Rezayi}. However, for $
\nu_{2}=1/6$, the CF liquid state is unexpectedly lower in energy than the
CDW state. There is one electron in each bubble for $\nu_{2}<1/6$ (The WC phase) and
two electrons for $\nu_{2}>1/6$ (the bubble phase). This means that a possible
metallic phase may exist in the crossover from the WC to the bubble phase. The RIQH
effect was previously explained as depinning and sliding of the WC and crystalizing
of the bubbles \cite{Cooper}. Unlike at the lowest LL, however, the lattice
constant at higher LLs does not change as one increases the LL filling but remains 
of the order of $R_{c}\gg l_{0}$. This implies the quantum fluctuations are hard
to depin the WC in the higher LLs \cite{Fogler}. In our picture, the WC {\it melts}
into the CF liquid and subsequently the bubbles containing $\tilde{M}=2$ electrons
form and are pinned as the filling increases. In fact, the feature of an isotropic 
metallic phase around $\nu_{2}=1/6$ can be identified in Fig.1 of Ref.\cite
{Cooper}. In addition, the $\nu_{2}=1/5$ state in our calculations is also 
slightly lower in energy than the corresponding CDW state while no FQHE was observed
in this region. This may be explained as the ground state energies of
compressible states (CDW and CF liquid) being lowered by disorder while
the incompressible FQHE state hardly being affected. For $n\geq 3$, 
the RIQH effects are even weaker to be identified 
in the experiment \cite{Cooper}. Our calculation also shows that it is very delicate. 
The CF cohesive energy is slightly higher than the CDW at $\nu_3=1/6$ (Fig. 3). 
However, the disorder should lower the CF ground state energy a little bit more 
than that of the CDW because of its gapless ground state. \newline
\indent In this work, we have presented a systematic computations of the
cohesive energies of the CF liquid, the Laughlin liquid and the CDW in the
lowest as well as in higher LLs. We find that the CF liquid and Laughlin
liquid generally dominate at the lowest and second LLs while the CDW
dominates at the higher LLs. The energy of the CF liquid at $\nu=4+1/6$ 
anomalously lower than that of the CDW may lead to the remarkable
phenomenon of RIQH. Since the difference of energies in this region is very slight
and the treatment of the screenings from the lower LLs is based 
on the Hartree-Fock approximation in the large $n$ limit, the results may not be 
completely decisive for $n=2,3$. Further numerical experiments are required 
to justify our conclusion. \newline
\indent One of the authors (SJY) thanks J. K. Jain for help. This work is
supported in part by the NSF of China. \newline

\begin{center}
FIGURES
\end{center}

Fig.1 The pair-distribution functions of Laughlin states at $n=0$ (the solid
line), $n=1$ (the dash-dotted line) and $n=2$ (the dotted line). (a) $\nu_{n}=1/3$.
(b) $\nu_{n}=1/5$.\newline
Fig.2 The pair-distribution functions of CF liquid states in the lowest LL
at $\nu=1/2$ (the solid line), $\nu=1/4$ (the dash-dotted line) and $\nu=1/6$
(the dotted).\newline
Fig.3 The cohesive energy per electron versus $\nu_{n}$ at the $n$-th LLs.
The bubble state and the stripe state are represented by the solid line and
dashed line, respectively. The triangle and square denote the Laughlin
state and the CF liquid state, respectively. The dotted line is a guide to
the eye. (a) $n=0$. (b) $n=1$. (c) $n=2$. (d) $n=3$.\newline

\begin{center}
TABLES\\[0pt]

\begin{tabular}{|c|c|c|c|}
\hline
$\nu_{n}$ & $E_{coh}^{B}$ & $E_{coh}^{UCDW}$ & $E_{coh}^{L}/E_{coh}^{CF}$ \\ 
\hline
1/2 & -0.1320 & -0.1120 & -0.1551 \\ 
1/3 & -0.1796 & -0.1469 & -0.2005 \\ 
1/4 & -0.1945 & -0.1574 & -0.2040 \\ 
1/5 & -0.1967 & -0.1584 & -0.2018 \\ 
1/6 & -0.1941 & -0.1547 & -0.1971 \\ 
1/8 & -0.1850 & -0.1466 & -0.1857 \\ \hline
\end{tabular}
\\[0pt]
\end{center}

Table 1. The cohesive energy per electron for $n=0$ LL. The last column are energies
for Laughlin states (odd-denominator fillings) or CF liquids (even-denominator fillings), 
respectively. The energy is in units of $e^{2}/\kappa l_{0}$.\newline

\begin{center}
\begin{tabular}{|c|c|c|c|}
\hline
$\nu_{n}$ & $E_{coh}^{B}$ & $E_{coh}^{UCDW}$ & $E_{coh}^{L}/E_{coh}^{CF}$ \\ 
\hline
1/2 & -0.0721 & -0.0687 & -0.0768 \\ 
1/3 & -0.0996 & -0.0918 & -0.1062 \\ 
1/4 & -0.1145 & -0.1014 & -0.1169 \\ 
1/5 & -0.1229 & -0.1055 & -0.1256 \\ 
1/6 & -0.1272 & -0.1069 & -0.1283 \\ 
1/8 & -0.1295 & -0.1063 & -0.1288 \\ \hline
\end{tabular}
\\[0pt]
\end{center}

Table 2. The same as Table 1 for $n=1$ LL.\newline

\begin{center}
\begin{tabular}{|c|c|c|c|}
\hline
$\nu_{n}$ & $E_{coh}^{B}$ & $E_{coh}^{UCDW}$ & $E_{coh}^{L}/E_{coh}^{CF}$ \\ 
\hline
1/2 & -0.0448 & -0.0456 & -0.0387 \\ 
1/3 & -0.0635 & -0.0598 & -0.0603 \\ 
1/4 & -0.0713 & -0.0659 & -0.0681 \\ 
1/5 & -0.0762 & -0.0690 & -0.0764 \\ 
1/6 & -0.0798 & -0.0707 & -0.0802 \\ 
1/8 & -0.0834 & -0.0719 & -0.0829 \\ \hline
\end{tabular}
\\[0pt]
\end{center}

Table 3. The same as Table 1 for $n=2$ LL.\newline

\begin{center}
\begin{tabular}{|c|c|c|c|}
\hline
$\nu_{n}$ & $E_{coh}^{B}$ & $E_{coh}^{UCDW}$ & $E_{coh}^{L}/E_{coh}^{CF}$ \\ 
\hline
1/2 & -0.0318 & -0.0329 & -0.0290 \\ 
1/3 & -0.0441 & -0.0435 & -0.0420 \\ 
1/4 & -0.0509 & -0.0480 & -0.0466 \\ 
1/5 & -0.0531 & -0.0502 & -0.0490 \\ 
1/6 & -0.0560 & -0.0514 & -0.0530 \\ 
1/8 & -0.0593 & -0.0523 & -0.0587 \\ \hline
\end{tabular}
\\[0pt]
\end{center}

Table 4. The same as Table 1 for $n=3$ LL.\newline

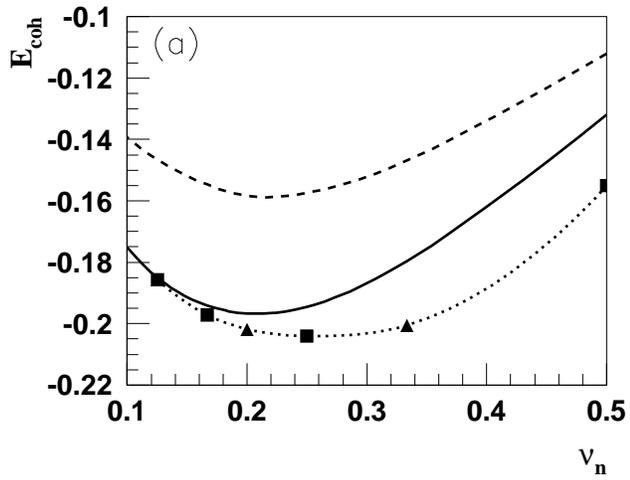
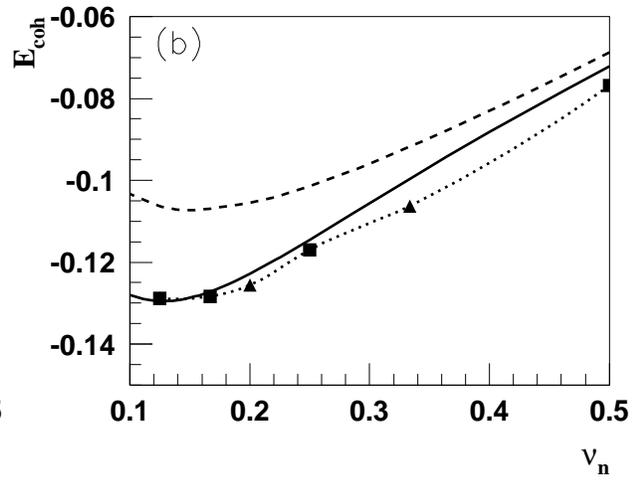
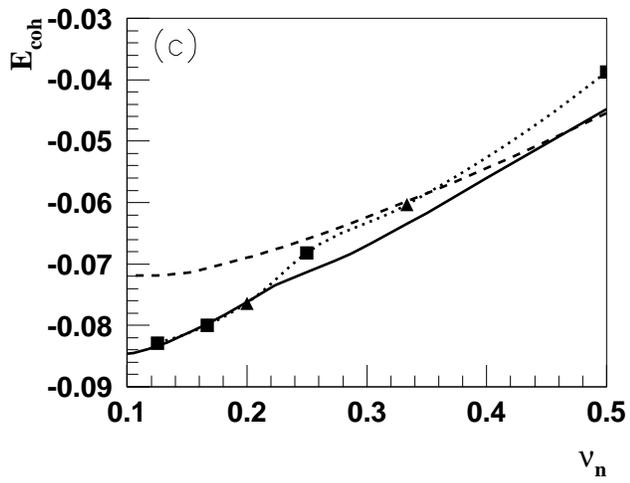
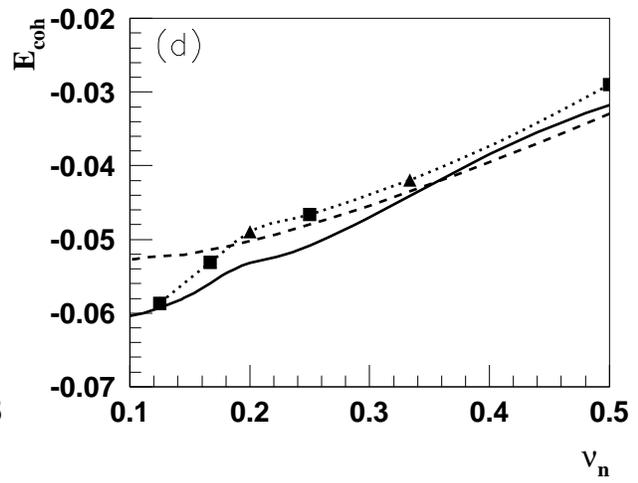
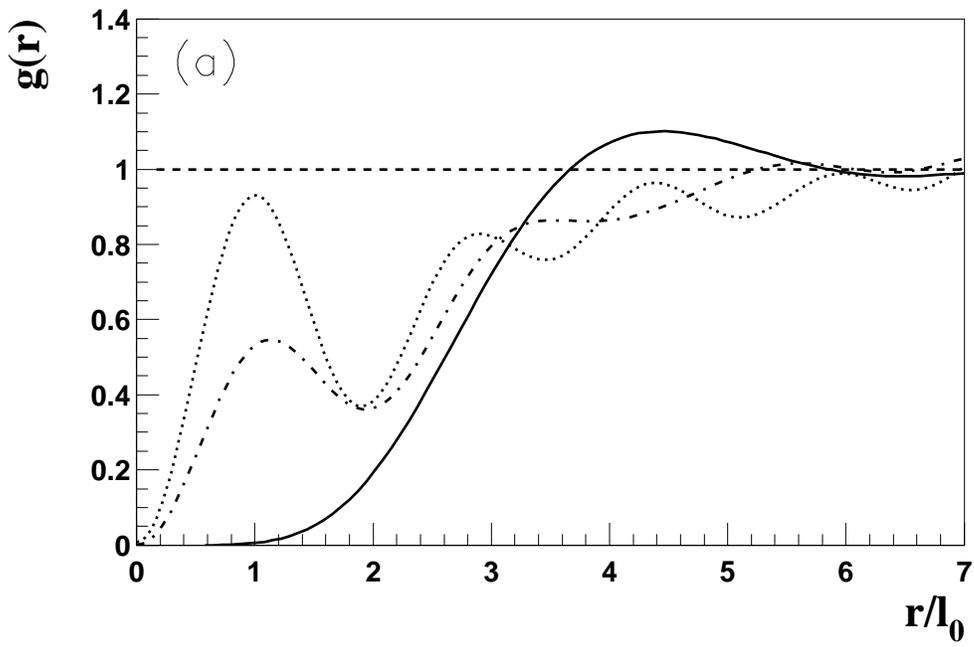
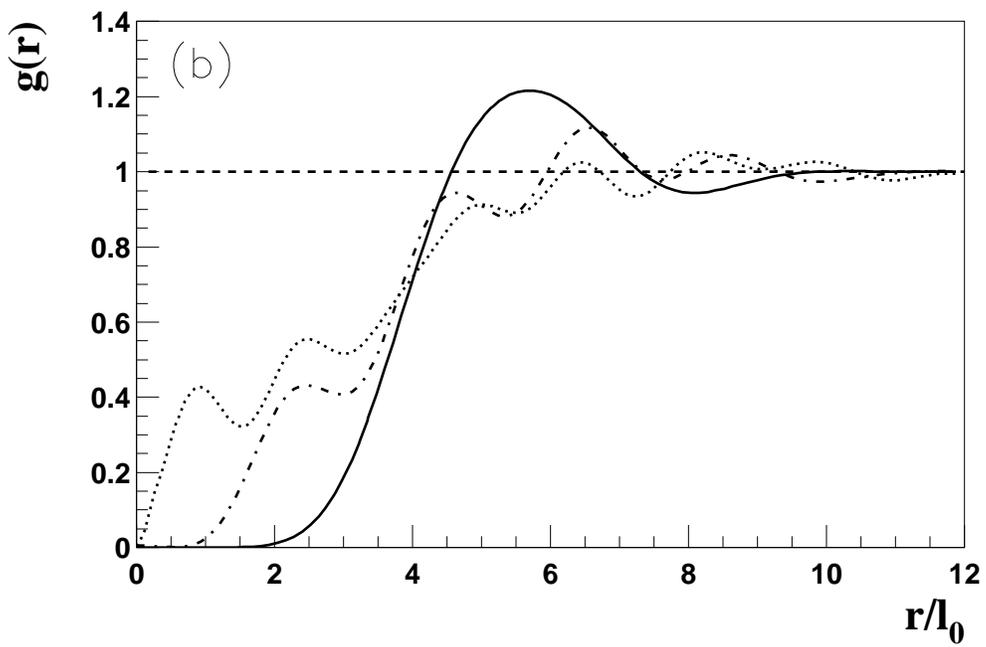
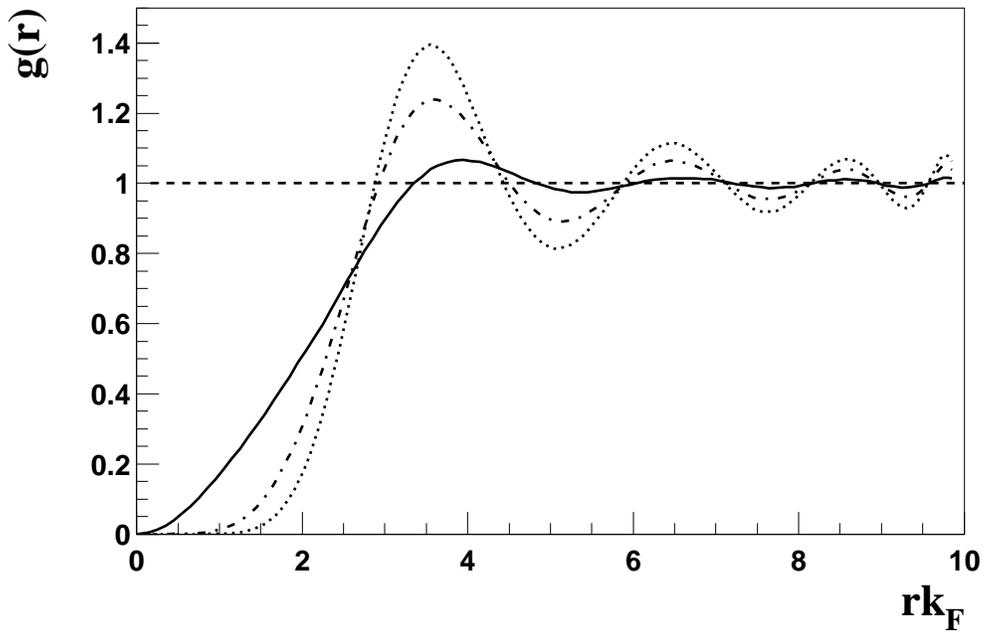
\end{document}